\documentclass[twocolumn,prb]{revtex4}

\usepackage{graphicx}    
\usepackage{dcolumn}     
\usepackage{bm}          
\usepackage{amsmath}

\begin{document}
\preprint{}

\title{
Zero-bias conductance peak splitting due to
multiband effect in tunneling spectroscopy}

\author{
Y. Tanuma,$^1$ K. Kuroki,$^2$
Y. Tanaka,$^3$ and S. Kashiwaya$^4$
}%
%
\affiliation{
$^1$Institute of Physics, Kanagawa University,
Rokkakubashi, Yokohama, 221-8686, Japan \\
$^2$Department of Applied Physics and
Chemistry, The University of Electro-Communications,
Chofu, Tokyo 182-8585, Japan \\
$^3$Department of Applied Physics, 
Nagoya University, Nagoya, 464-8603, Japan \\
$^4$National Institute of Advanced Industrial Science
and Technology, Tsukuba, 305-8568, Japan
}
%
\date{\today}

\begin{abstract}
We study how the multiplicity of the Fermi surface 
affects the zero-bias peak in conductance spectra of tunneling spectroscopy.
As case studies, we consider models for organic superconductors
$\kappa$-(BEDT-TTF)$_2$Cu(NCS)$_2$ and (TMTSF)$_2$ClO$_4$.
We find that multiplicity of the Fermi surfaces can lead to a
splitting of the zero-bias conductance peak (ZBCP). 
We propose that the presence/absence of the ZBCP splitting 
is used as a probe to distinguish the pairing symmetry in 
$\kappa$-(BEDT-TTF)$_2$Cu(NCS)$_2$.
\end{abstract}

\maketitle

\section{Introduction}
An unambiguous determination of pairing symmetry
in unconventional superconductors is crucial
to understand the pairing mechanism of superconductivity.
Strong evidences suggesting $d_{x^2-y^2}$-wave pairing symmetry 
in the high-$T_c$ cuprates have been provided 
using several phase-sensitive probes
\cite{SR95,Harlin,Tsuei}
including tunneling spectroscopy
via Andreev surface bound states (ABS's).
\cite{Buch,Hu,TK95,FRS97}
The tunneling spectroscopy via ABS's enables us
to detect the sign change
in the pair potential as well as its nodal structure.
\cite{TK95,KT00,Lofwander}
This state,
which originates from the interference effect in the
effective pair potential of
the $d_{x^{2}-y^{2}}$-wave symmetry
through reflection at a surface or an interface,  
have significant influences on several charge transport 
properties \cite{c1,c2,c3}.
The existence of ABS's,
which manifests itself as a distinct conductance peak at zero-bias
in the tunneling spectrum (zero-bias conductance peak, referred to as ZBCP),
has been actually observed
not only in the high-$T_c$ cuprates
\cite{exp1}
but also in ruthenates \cite{Laube,Mao1},
heavy fermion systems \cite{Walti},
and more recently $\mathrm{MgCNi}_3$. \cite{Mao2}
In this context, it is of great interest to investigate
whether the ZBCP due to the ABS's can be observed
in organic superconductors \cite{IYS} such as
$\kappa$-(BEDT-TTF)$_2X$ 
and $\textrm{(TMTSF)}_2X$ \cite{organic}.
\par
The tunneling spectroscopy via ABS's can be used to 
determine the pairing symmetry
if one can prepare well-treated surfaces
with arbitrary orientations in the superconducting plane.
For high-$T_c$ cuprates, which has a 
$d_{x^2-y^2}$ pair potential, 
it is theoretically shown that
the ZBCP should be observed most prominently for (110) 
surfaces or interfaces.
Moreover, it has been clarified that
the ZBCP may be observed due to atomic-scale roughness 
even in the (100) surfaces.
\cite{FRS97,Tanu1,Tanu2,Samo}
In fact, Iguchi et al. \cite{Iguchi} have measured the ZBCP
for Ag/YBCO ramp-edge junctions with various orientations,
where the injection direction 
varies continuously from (100) to (110) interfaces.
The height of the ZBCP has shown to vary according as the misorientation
angle from $a$-axis  within the plane.
\par
As regards
organic superconductors such as $\kappa$-(BEDT-TTF)$_2X$,
the pairing symmetry of the pair potentials
still remains to be a controversial problem.
It has indeed become an issue of great interest 
whether $\kappa$-(BEDT-TTF)$_2X$
has a $d$-wave pair potential similar to high-$T_c$ cuprates.
There is now a body of accumulating experimental evidences
suggesting that $\kappa$-(BEDT-TTF)$_2X$
have anisotropy in the pair potential.
\cite{Mayaffre,Soto,Kanoda96,Nakazawa,Carrington,Pinteric}
Earlier theories support 
$d_{x^2-y^2}$-wave pairing, \cite{Schma,KKKM,KA99}
while a recent thermal conductivity measurement 
suggests $d_{xy}$-wave pairing.
\cite{Izawa3}
Concerning this issue,
two of the present authors have theoretically shown 
that a $d_{xy}$-like pairing may slightly dominate
over $d_{x^2-y^2}$ pairing
when the dimerization of the BEDT-TTF molecules
is not so strong.
\cite{Kuro02}
According to previous studies,
\cite{Hu,TK95,FRS97}
if the pairing symmetry of $\kappa$-(BEDT-TTF)$_2X$
is $d$-wave, ABS is expected to be created at surfaces 
for arbitrary injection orientations.
However, a scanning tunneling
microscopy (STM) experiment for $\kappa$-(BEDT-TTF)$_2X$
by Arai et al.\cite{Arai}
showed the absence of 
ZBCP for arbitrary injection angle from the $c$-axis
in the $bc$-plane,
which is in contrast with the case of the high-$T_c$ cuprates.
%
The presence/absence of the ZBCP
of $d$-wave superconductors is sensitive to several factors:
(i) roughness effect of surfaces or interfaces,
(ii) random impurity scattering effect near the interfaces, 
(iii) the shape of the Fermi surface,
and (iv) the degradation of surfaces.
The disappearance of the ZBCP in $d$-wave superconductors
due to reason (i) has been studied previously\cite{Tanu1,Tanu2}. 
Depending on the shape of the Fermi surface and 
the geometry of the surface, the atomic-size wave nature of the 
zero energy ABS's (ZES), $i.e.$,
the oscillatory behavior of the wave function 
of ZES induces an interference effect which locally destroys the ZBCP. 
In fact, it is by no means easy to make well-oriented
cleavage surfaces in organic materials, so this point may be 
important.
As regards point (ii), 
Asano $et$ $al.$ \cite{asano} have shown,
both from analytical and 
numerical calculation  beyond quasiclassical approximations 
\cite{random},  
that impurity scattering near the interface
in the high-$T_{C}$ cuprates  
can induce a splitting  or a 
disappearance of the ZBCP.
\par
%
As for point (iii), 
we have recently studied the disappearance of ZBCP
due to the warping of
quasi-1D Fermi surface as in $\mathrm{(TMTSF)}_2X$.
\cite{Tanu3}
The results indicate that the ABS's are
sensitive to the shape of the Fermi surface.
However, most of the theoretical studies on tunneling spectroscopy
via ABS's up to date have been performed 
for single band systems.
It has not been clarified how the multiplicity of the Fermi surface
influences the ZBCP.
\par
Motivated by this point, here we investigate the
surface density of states in systems having multiple Fermi surfaces,
where we focus on two organic superconductors as case studies, namely,  
$\kappa$-(BEDT-TTF)$_2$Cu(NCS)$_2$ and (TMTSF)$_2$ClO$_4$.
The Fermi surface of $\kappa$-(BEDT-TTF)$_2$Cu(NCS)$_2$,
which has been determined by Shubnikov-de Haas experiment,
\cite{Oshima}
consists of two portions separated by small gaps.
The Fermi surface of (TMTSF)$_2$ClO$_4$ is also separated by 
a small gap, which is due to anion ordering.
In this paper, we extend our previous studies
\cite{Tanu3}
on anisotropic triangular lattice
by taking into account these multiplicity of the Fermi surface.
\par
The organization of the paper is as follows.
The formulation of calculating 
the tunneling spectrum on anisotropic triangular lattice
is presented in Sec.~\ref{sec:02}.
In Sec.~\ref{sec:03},
results of the numerical calculations 
are discussed in detail.
Finally,
we summarize the paper in Sec.~\ref{sec:04}.
\par
%
\section{Formulation}
\label{sec:02}
%
\begin{figure}[htb]
\begin{center}
\scalebox{0.40}{
\includegraphics{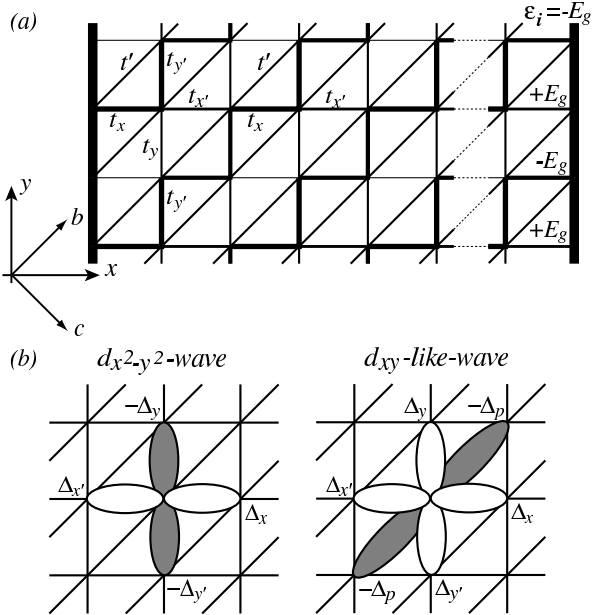}}
\caption{
(a) Schematic of (100) surface in $xy$-plane
with next-nearest neighbor hopping $t^{\prime}$.
(b) Cooper pairs in real space for $d_{x^2-y^2}$
and $d_{xy}$-like-wave pairings.
\label{fig:01}}
\end{center}
\end{figure}
%
In the present study,
we start from an
extended Hubbard model given by
\begin{align}
\label{Hami01}
    {\mathcal H} =
 &- \sum_{\bm{i},\bm{j},\sigma}t_{\bm{ij}}
  c^{\dagger}_{\bm{i},\sigma} c_{\bm{j},\sigma}
 - \frac{V}{2}\sum_{\bm{i},\bm{j},
      \sigma,\sigma^{\prime}}
      c^{\dagger}_{\bm{i},\sigma}
      c^{\dagger}_{\bm{j},\sigma^{\prime}}
      c_{\bm{j},\sigma^{\prime}}
      c_{\bm{i},\sigma},
\nonumber \\
 &+ \sum_{\bm{i},\sigma}(\varepsilon_{\bm{i}} -\mu)
  c^{\dagger}_{\bm{i},\sigma} c_{\bm{i},\sigma},
\end{align}
where $c_{\bm{i},\sigma}^{\dagger}$
creates a hole with spin $\sigma=\uparrow,\downarrow$
at site $\bm{i}=(i_{x},i_{y})$.
As a model for $\kappa$-(BEDT-TTF)$_2$Cu(NCS)$_2$, 
each site corresponds to BEDT-TTF molecule dimmers. 
We consider five kinds of hopping integrals,
$t_{x}(=t)$, $t_{x'}$, $t_{y}$, $t_{y'}$,
and $t^{\prime}$ in the $xy$ plane
on the anisotropic triangular lattice
as shown in Fig.~\ref{fig:01}(a).
In order to reproduce the shape of Fermi surface
for $\kappa$-(BEDT-TTF)$_2$Cu(NCS)$_2$ and
$\mathrm{(TMTSF)}_2\mathrm{ClO}_4$,
we adopt the values of
(i) $t^{\prime}=0.8t$, $t_{y'}=t_{x}$, $t_{x'}=t_{y}$,
\cite{Oshima}
and (ii) $t_{y}=0.1t$, $t_{x'}=t_{x}$, $t_{y'}=t_{y}$,
\cite{Ducasse,Grant,Yamaji}
respectively.
Two subchains in the $x$ direction
alternatively have the site energy
$\varepsilon_{\bm{i}}=+E_{g},-E_{g},+E_{g}, \dots$
in the $y$ direction.
\cite{Shimahara,Yoshino}
The chemical potential $\mu$
is determined such that the band in $\kappa$-(BEDT-TTF)$_2$Cu(NCS)$_2$
[$\mathrm{(TMTSF)}_2\mathrm{ClO}_4$]
is half-filled [quarter-filled].
The effective attraction $V$ is assumed to act 
on a pair of electrons.
\par
By solving the mean-field equation
for a unit cell with $N_{\rm L}(=500)$ sites
in the $x$ direction
and two sites in the $y$ direction, 
we obtain the eigenenergy $E_{\nu}$.
In terms of the eigenenergy $E_{\nu}$ and the 
wave functions $u^{\nu}_{\bm{i}}$, $v^{\nu}_{\bm{i}}$,
the Bogoliubov-de Gennes equation for the (100) surface
in the $xy$ plane is given by
\begin{eqnarray}
\label{eq:HF}
    \sum_{\bm{j}}
     \left (
      \begin{array}{cc}
    H_{\bm{ij}} &
    F_{\bm{ij}} \\
    F^{*}_{\bm{ij}} &
   -H_{\bm{ij}}^{*} \\
      \end{array}
     \right )
     \left (
      \begin{array}{c}
            u^{\nu}_{\bm{j}} \\
            v^{\nu}_{\bm{j}} \\
      \end{array}
     \right )
  =E_{\nu}
     \left (
      \begin{array}{c}
            u^{\nu}_{\bm{i}} \\
            v^{\nu}_{\bm{i}} \\
      \end{array}
     \right ),
\end{eqnarray}
with
\begin{align}
H_{ij}(k_{y}) =
        &-t_{x}\eta_{+}\delta_{j_x,i_x+1}
         -t_{x'}\eta_{-}\delta_{j_x,i_x+1}
    \nonumber \\
       & -t_{y}e^{-2\mathrm{i}k_{y}\delta_{i_{y},2}}
         \eta_{+} \delta_{j_y,i_y+1}
    \nonumber \\
       & -t_{y'}e^{-2\mathrm{i}k_{y}\delta_{i_{y},2}}
         \eta_{-} \delta_{j_y,i_y+1}
    \nonumber \\
  &     -t^{\prime}e^{-2\mathrm{i}k_{y}\delta_{i_{y},2}}
         \delta_{j_{x},i_{x}+1}\delta_{j_{y},i_{y}+1}
\nonumber \\
       & -t_{x}\eta_{-}\delta_{j_x,i_x-1}
         -t_{x'}\eta_{+}\delta_{j_x,i_x-1}
    \nonumber \\
       & -t_{y}e^{2\mathrm{i}k_{y}\delta_{i_{y},1}}
         \eta_{-} \delta_{j_y,i_y-1}
    \nonumber \\
       & -t_{y'}e^{2\mathrm{i}k_{y}\delta_{i_{y},1}}
         \eta_{+} \delta_{j_y,i_y-1}
    \nonumber \\
       & -t^{\prime}e^{2\mathrm{i}k_{y}\delta_{i_{y},1}}
         \delta_{j_{x},i_{x}-1}\delta_{j_{y},i_{y}-1}
\nonumber \\
       &+\left \{ (-1)^{i_y+1}E_{g}
        -\mu \right \} \delta_{i_x,j_x}\delta_{i_y,j_y},
\end{align}
where we define $\eta_{+}=\frac{1}{2}\{ 1+(-1)^{i_x+i_y} \}$
and $\eta_{-}=\frac{1}{2}\{ 1-(-1)^{i_x+i_y} \}$.
\par
%
\begin{figure}[htb]
\begin{center}
\scalebox{0.45}{
\includegraphics{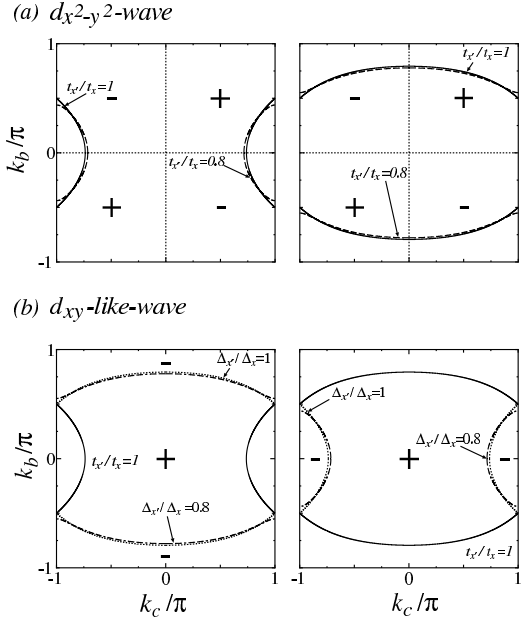}}
\caption{
The Fermi surface and $d$-wave pairings.
(a) $d_{x^2-y^2}$ and (b) $d_{xy}$-like-wave.
\label{fig:02}}
\end{center}
\end{figure}
%
As for plausible pairing symmetries
in $\kappa$-(BEDT-TTF)$_2$Cu(NCS)$_2$,
we consider $d_{x^2-y^2}$-wave pairing given by,
\begin{align}
F_{ij}(k_{y}) &=
         \Delta_{x}\eta_{+}\delta_{j_x,i_x+1}
        +\Delta_{x'}\eta_{-}\delta_{j_x,i_x+1}
    \nonumber \\
       &-\Delta_{y}e^{-2\mathrm{i}k_{y}\delta_{i_{y},2}}
         \eta_{+} \delta_{j_y,i_y+1}
    \nonumber \\
       &-\Delta_{y'}e^{-2\mathrm{i}k_{y}\delta_{i_{y},2}}
         \eta_{-} \delta_{j_y,i_y+1}
    \nonumber \\
       &+\Delta_{x}\eta_{-}\delta_{j_x,i_x-1}
        -\Delta_{x'}\eta_{+}\delta_{j_x,i_x-1}
    \nonumber \\
       &-\Delta_{y}e^{2\mathrm{i}k_{y}\delta_{i_{y},1}}
         \eta_{-} \delta_{j_y,i_y-1}
    \nonumber \\
       &-\Delta_{y'}e^{2\mathrm{i}k_{y}\delta_{i_{y},1}}
         \eta_{+} \delta_{j_y,i_y-1}
\end{align}
and  $d_{xy}$-like pairing given by 
\begin{align}
F_{ij}(k_{y}) &=
         \Delta_{x}\eta_{+}\delta_{j_x,i_x+1}
        +\Delta_{x'}\eta_{-}\delta_{j_x,i_x+1}
    \nonumber \\
       &+\Delta_{y}e^{-2\mathrm{i}k_{y}\delta_{i_{y},2}}
         \eta_{+} \delta_{j_y,i_y+1}
    \nonumber \\
       &+\Delta_{y'}e^{-2\mathrm{i}k_{y}\delta_{i_{y},2}}
         \eta_{-} \delta_{j_y,i_y+1}
    \nonumber \\
  &     -\alpha \Delta_{p}e^{-2\mathrm{i}k_{y}\delta_{i_{y},2}}
         \delta_{j_{x},i_{x}+1}\delta_{j_{y},i_{y}+1}
    \nonumber \\
       &+\Delta_{x}\eta_{-}\delta_{j_x,i_x-1}
        +\Delta_{x'}\eta_{+}\delta_{j_x,i_x-1}
    \nonumber \\
       &+\Delta_{y}e^{2\mathrm{i}k_{y}\delta_{i_{y},1}}
         \eta_{-} \delta_{j_y,i_y-1}
    \nonumber \\
       &+\Delta_{y'}e^{2\mathrm{i}k_{y}\delta_{i_{y},1}}
         \eta_{+} \delta_{j_y,i_y-1}
    \nonumber \\
  &     -\alpha \Delta_{p}e^{2\mathrm{i}k_{y}\delta_{i_{y},1}}
         \delta_{j_{x},i_{x}-1}\delta_{j_{y},i_{y}-1},
\end{align}
with $\alpha=0.8t$ in accord with Ref.~\onlinecite{Kuro02}.
The pairing in real space is shown in Fig.~\ref{fig:01}(b).
Here, we select $\Delta_{x'}=\Delta_{y}$ and
$\Delta_{x}=\Delta_{y'}=\Delta_{p}=\Delta_0$,
where $\Delta_0$ is a bulk value.
\par
%
\begin{figure}[htb]
\begin{center}
\scalebox{0.40}{
\includegraphics{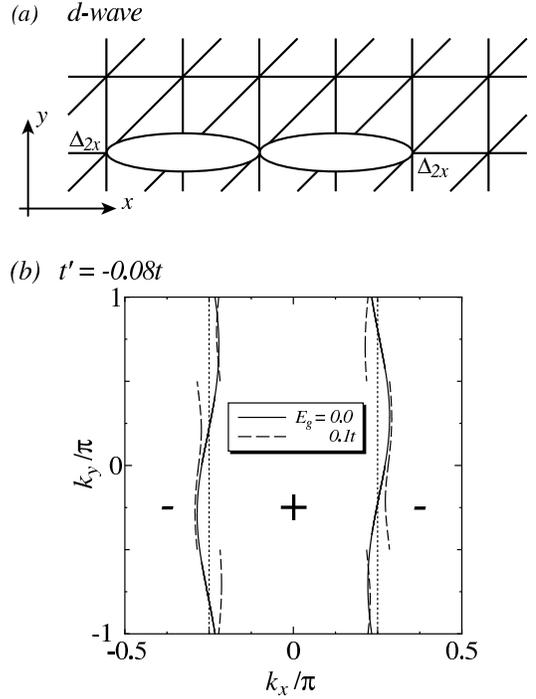}}
\caption{
(a) Cooper pairs with $d$-wave symmetry
separated by two lattice space and
(b) quasi-1D Fermi surface
in $t_y=0.1t_x$ and $t^{\prime}=-0.08t_x$.
\label{fig:03}}
\end{center}
\end{figure}
%
The upper and lower panels of Fig.~\ref{fig:02}
show the $d_{x^2-y^2}$ and $d_{xy}$-like pair potentials 
in momentum space along with the Fermi surfaces.
For $t_{x'}=t_{x}$, the Fermi surface is continuously
connected.
In the actual $\kappa$-(BEDT-TTF)$_2$Cu(NCS)$_2$, however, 
BEDT-TTF dimmers are further dimerized so that $t_{x'} \neq t_{x}$,
which leads to a splitting of the Fermi surface 
at around $(k_c,k_b)=(\pm \pi,\pm \pi/2)$.
%
%
\par
%
\begin{figure}[htb]
\begin{center}
\scalebox{0.50}{
\includegraphics{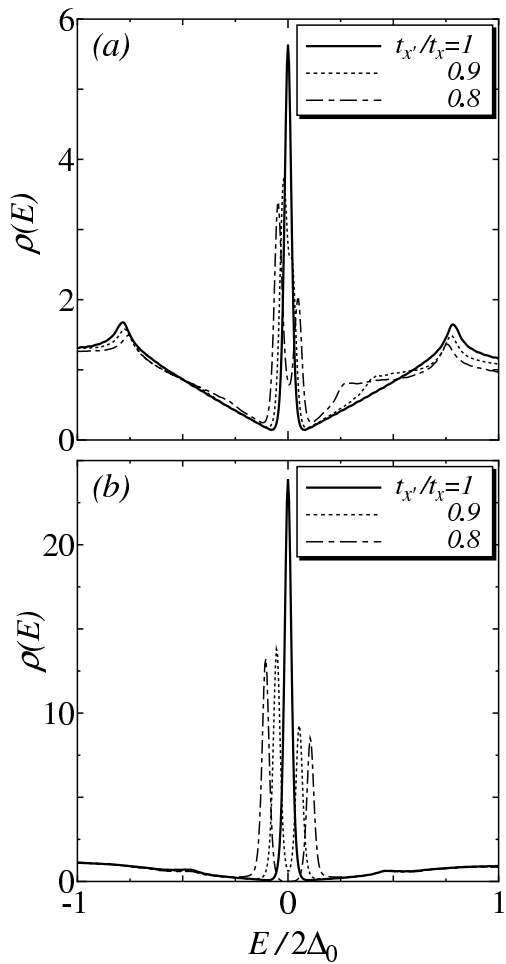}}
\caption{
Tunneling spectrum
for (a) $d_{x^2-y^2}$ and (b) $d_{xy}$-like-waves
fixed in $\Delta_{x'}=\Delta_{x}$.
\label{fig:04}}
\end{center}
\end{figure}
%
%
In $\mbox{(TMTSF)}_{2}\mbox{ClO}_{4}$,
the orientational order of the anions $\mbox{ClO}_{4}$
doubles the unit cell,
leading again to a splitting of the Fermi surface. \cite{Pouget83}
Although the pairing symmetry for $\mbox{(TMTSF)}_{2}X$
remains to be undetermined, 
here we assume singlet $d$-wave pairing as an example 
in which the multiband effect is prominent.
In this case $d$-wave
is a pairing separated by two lattice spacings given by
\begin{align}
F_{ij} =
         \Delta_{2x}\delta_{j_x,i_x+2}
        +\Delta_{2x}\delta_{j_x,i_x-2}.
\end{align}
%
%
The anion potential $E_g$ is estimated
from experimental measurement of
angle dependent magnetoresistance.
\cite{Yoshino}
%
\par
In order to compare our theory with
STM experiments,
we assume that the STM tip is metallic
with a constant density of states, 
and that tunneling occurs only
to the site nearest to the tip.
This has been shown to be valid 
through the study of tunneling conductance of unconventional
superconductors \cite{TK95}. 
The tunneling conductance spectrum is then given
at low temperatures by 
the normalized surface density of states \cite{TK95},
\begin{align}
   \rho(E) &=
   \frac{
   \displaystyle{
   \int ^{\infty}_{-\infty}{\rm d}\omega \rho_{\rm S}(\omega)
{\rm sech}^{2}\left ( \frac{\omega + E}{2k_{\rm B}T} \right )}}
   {
   \displaystyle{
   \int ^{\infty}_{-\infty}{\rm d}\omega \rho_{{\rm N}}(\omega)
{\rm sech}^{2}\left ( \frac{\omega - 2\Delta_{0}}{2k_{\rm B}T} \right )}},
\\
\rho_{\rm S}(\omega) &=
\sum_{k_{b},\nu} \left [|u^{\nu}_{1}|^{2}
\delta(\omega - E_{\nu})
+ |v^{\nu}_{1}|^{2}
\delta(\omega + E_{\nu}) \right ].
\end{align}
Here $\rho_{\rm S}(\omega)$ denotes the 
surface density of states 
for the superconducting state while $\rho_{\rm N}(\omega)$
the bulk density of states in the normal state.
\par
%
\begin{figure}[htb]
\begin{center}
\scalebox{0.45}{
\includegraphics{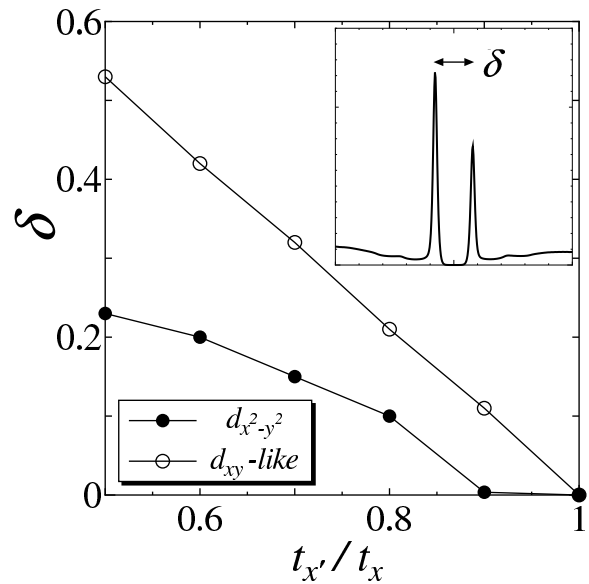}}
\caption{
The $t_{x'}/t_x$ v.s.
the zero-energy peak splitting width $\delta$.
\label{fig:05}}
\end{center}
\end{figure}
\section{Results of in-plane tunneling spectrum}
\label{sec:03}
%
In this section, we present the calculation results.
First, let us focus on the model for $\kappa$-(BEDT-TTF)$_2X$.
We examine the case of the tunneling spectrum
at (100) surface on the $xy$ plane
as shown in Fig.~\ref{fig:01}.
As seen in Fig.~\ref{fig:04},
in the case of $t_{x'}=t_{x}$,
where the Fermi surface is elliptical but 
continuous, there exists a distinct peak at zero energy, 
which resembles those obtained in 
previous theories assuming round shape Fermi surface.
The ZEP arises because 
incident and reflected (including oblique incidence)
quasiparticles normal to the surface 
feel opposite signs of the pair potential, which 
results in a formation of the ABS.
If we turn on the multiband effect 
by letting  $t_{x'}\neq t_{x}$, the ZEP is found to 
split into two.
This is reminiscent of 
the ZEP splitting originating from broken
time reversal symmetry states.\cite{FRS97,matsumoto2,Tanu5,K95} 
\par
%
\begin{figure}[htb]
\begin{center}
\scalebox{0.45}{
\includegraphics{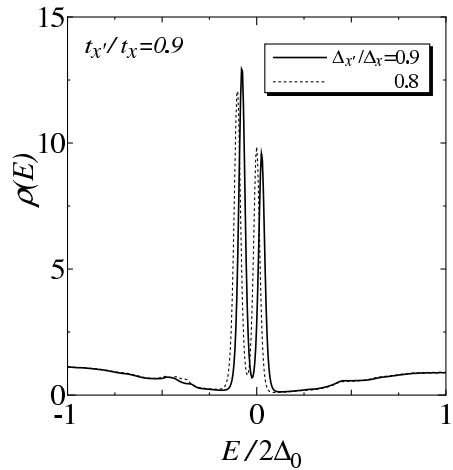}}
\caption{
Tunneling spectrum
for variour $\Delta_{x'}/\Delta_{x}$
in $t_{x'}/t_{x}=0.9$.
\label{fig:06}}
\end{center}
\end{figure}
%
We have further studied the $t_{x'}/t_{x}$ dependence of
the ZEP splitting.
In Fig.~\ref{fig:05},
the width of the ZEP splitting $\delta$
is plotted as functions of $t_{x'}/t_{x}$
for $d_{x^2-y^2}$ and $d_{xy}$-like pairings.
$\delta$ for the $d_{xy}$-like pair potential
is almost proportional to $t_{x'}/t_{x}$,
and larger than that for $d_{x^2-y^2}$.
In the regime of $t_{x'}/t_{x}>0.9$, in particular, 
we see no splitting for the $d_{x^2-y^2}$ pairing.
Since $t_x'/t_x$ is estimated to be $\sim 0.9$,
\cite{Komatsu} 
we may be able to distinguish 
between $d_{x^2-y^2}$ and $d_{xy}$-like pairings
through the presence/absence of ZEP splitting.
\par
We have also performed similar calculation 
by letting $\Delta_{x'}/\Delta_{x}$ deviate from unity,
which should be the case when $t_x$ deviates from $t_x$.
The results are plotted in Fig.~\ref{fig:06}
for various $\Delta_{x'}/\Delta_{x}$ with 
$t_{x'}/t_{x}$ fixed at $0.9$.
In this case, we observe an overall shift of the splitted ZEP,
while the magnitude of the splitting remains unchanged.
\par
\begin{figure}[htb]
\begin{center}
\scalebox{0.55}{
\includegraphics{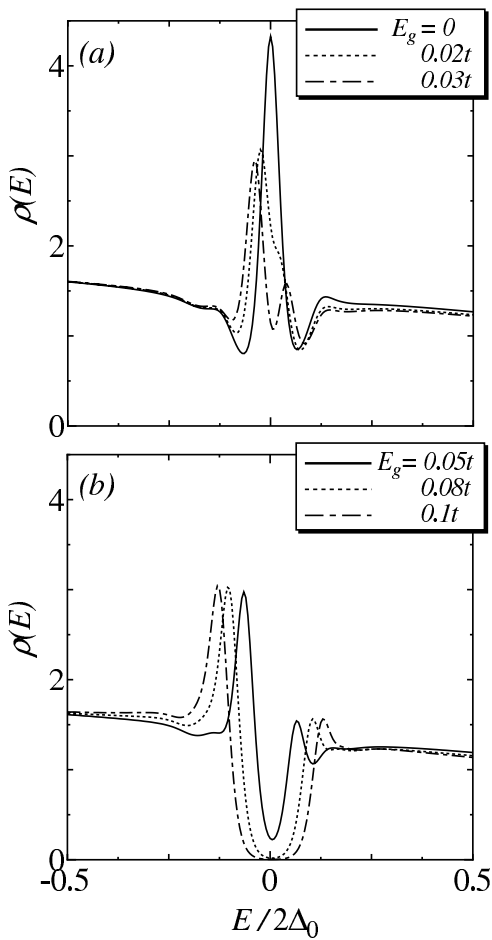}}
\caption{
Tunneling spectrum in $t_{y}=0.1t_{x}$ and $t^{\prime}=
-0.08t_{x}$ with anion potential $E_{g}$.
\label{fig:07}}
\end{center}
\end{figure}

Let us now move on to the model for $\mbox{(TMTSF)}_{2}\mbox{ClO}_{4}$.
In this model, as the Fermi surface becomes asymmetric
with respect to $k_x\rightarrow -k_x$ transformation,
some injected and reflected quasiparticles
feel different signs and the ZEP appears 
in the tunneling spectrum \cite{Tanu3}.
When $E_g$ is turned on,
a minigap opens at $k_{y}=\pm \pi/4$.
This effect again leads to ZEP splitting, of which the magnitude increases
as $E_g$ is increased.
[Fig.~\ref{fig:07}]
\par
Although it is by no means easy to pinpoint the 
origin of the ZEP splitting analytically, 
it can be qualitatively explained as follows. 
For single band models, 
the ZEP appears due to the sign change of the pair potential 
felt by quasiparticles at the interface. 
In multiband systems, injected and reflected quasiparticles 
have  different band indices, so that an additional 
phase factor is expected to appear due to interband scattering. 
This additional phase factor induces the ZEP splitting as in the 
case of pair potentials with broken time reversal symmetry. 
%
\section{Conclusions}
\label{sec:04}
%
To summarize, 
we have investigated the multiband effect on tunneling spectroscopy.
As case studies,
we have focused on models for 
$\kappa$-(BEDT-TTF)$_2$Cu(NCS)$_2$ 
and $\mathrm{(TMTSF)}_2\mathrm{ClO}_4$.
We find that the multiplicity of the Fermi surface can 
lead to a splitting of the ZEP. 
As regards $\kappa$-(BEDT-TTF)$_2$Cu(NCS)$_2$,  
since $t_x'/t_x$ is estimated to be $\sim 0.9$,
\cite{Komatsu} 
we can distinguish 
between $d_{x^2-y^2}$ and $d_{xy}$-like pairings
through the presence/absence of ZEP splitting. 
\par
As mentioned in the Introduction, however, 
a scanning tunneling measurement 
\cite{Arai} actually find no ZBCP in the tunneling spectrum of 
$\kappa$-(BEDT-TTF)$_2$Cu(NCS)$_2$.
%
%
Since many experiments suggest the existence of nodes in the 
pair potential in $\kappa$-(BEDT-TTF)$_2X$, we believe that 
the absence of ZBCP is not because the pairing symmetry is a simple $s$-wave,
but because of the roughness of the surface or
the random scattering effect by impurities near the 
interface\cite{random}, namely, point (i) or (ii) mentioned in the 
Introduction. 
As regards the roughness of the surface,
we believe it is necessary to study tunneling spectroscopy 
of $\kappa$-(BEDT-TTF)$_2$Cu(NCS)$_2$ in the presence of 
atomic scale roughness as done by 
Tanuma $et$ $al.$ on a lattice model.\cite{Tanu1,Tanu2} 
As for the issue of random scattering effect by impurities, 
Asano $et$ $al.$ \cite{asano} have shown,
both from analytical and 
numerical calculation  beyond quasiclassical approximations,  
that impurity scattering near the interface in the high-$T_{C}$ cuprates  
can induce a splitting  or a 
disappearance of the ZBCP.
From this viewpoint, 
it would also be interesting to study the impurity scattering 
effect in $\kappa$-(BEDT-TTF)$_2$Cu(NCS)$_2$.
\par
In order to clearly determine the pairing symmetry, 
other complementary probes should also be used.
Recently, we have shown that
magnetotunneling spectroscopy is a promising method to 
identify the detailed paring symmetry of the unconventional 
superconductors \cite{Tanu3,Tanu4,magnet}. 
It would also be interesting to apply this probe to 
$\kappa$-(BEDT-TTF)$_2$Cu(NCS)$_2$.
\par
It is well known that ABS's have serious influence on the 
Josephson current. There are many works on 
Josephson effect in unconventional superconductors 
both from theoretical and experimental 
view point \cite{Josephson}. 
It is also a future problem to study Josephson effect
in $\kappa$-(BEDT-TTF)$_2$Cu(NCS)$_2$ 
and $\mathrm{(TMTSF)}_2\mathrm{ClO}_4$.

\section*{Acknowledgments}
The computations have been performed
at the Supercomputer Center of
Yukawa Institute for Theoretical Physics,
Kyoto University.

\end{document}